\documentclass[12pt]{article}

\usepackage[cp1251]{inputenc}

\usepackage{amsfonts}
 \usepackage{amsthm}
 \usepackage{amsmath}
 \usepackage{amscd}
 \usepackage{amssymb}

\usepackage[english]{babel}
\usepackage{enumitem}

 \usepackage{graphics}
\usepackage{graphicx}

 \numberwithin{equation}{section}
 \newtheorem{thm}{Theorem}
 
 \newtheorem{prop}{Proposition}
 \theoremstyle{definition}

\title{Non-integrability of a three dimensional generalized H\'{e}non-Heiles system}
\author{Ognyan Christov\\
Faculty of Mathematics and Informatics, Sofia University, \\
5 J. Bouchier blvd., 1164 Sofia, Bulgaria}
\date{}

\begin{document}

\maketitle

\begin{abstract}
\noindent
In recent paper Fakkousy et al. show that the 3D
H\'{e}non-Heiles system with Hamiltonian $ H = \frac{1}{2} (p_1 ^2
+ p_2 ^2 + p_3 ^2) +\frac{1}{2} (A q_1 ^2 + C q_2 ^2 + B q_3 ^2) +
(\alpha q_1 ^2 + \gamma q_2 ^2)q_3 + \frac{\beta}{3}q_3 ^3 $ is
integrable in  sense of Liouville when $\alpha = \gamma,
\frac{\alpha}{\beta} = 1, A = B = C$; or $\alpha = \gamma,
\frac{\alpha}{\beta} = \frac{1}{6}, A = C$, $B$-arbitrary; or
$\alpha = \gamma, \frac{\alpha}{\beta} = \frac{1}{16}, A = C,
\frac{A}{B} = \frac{1}{16}$ (and of course, when $\alpha=\gamma=0$,
in which case the Hamiltonian is separable). It is known that the second
case remains integrable  for $A, C, B$ arbitrary. Using Morales-Ramis theory, we prove
that there are no other cases of integrability for this system.
\end{abstract}


\section{Introduction}
\label{intro}

The generalized two-degrees of freedom H\'{e}non-Heiles system (H-H2)
is defined by the Hamiltonian
\begin{equation}
\label{hh}
H = \frac{1}{2} (p_1^2 + p_2^2) + \frac{1}{2}(A q_1^2 + B q_2^2) + \alpha q_1^2 q_2 + \frac{\beta}{3} q_2^3.
\end{equation}
The original H\'{e}non-Heiles Hamiltonian \cite{HH} agrees with (\ref{hh}) when $A = B = \alpha = 1, \beta = -1$.
This famous system appears as an important model in a large set of problems in Celestial, Statistical and Quantum Mechanics,
see for example the references in \cite{Wenlei,Fer1,Fer2,Eilbeck,Kostov,Zeng,Fakkousy}.

It is well known that the integrable cases of the above Hamiltonian are:

\vspace{2ex}

(0) $\alpha = 0$ -- the Hamiltonian is separable;

(1) $\beta = \alpha$ and $ A = B$;

(2) $\beta = 6 \alpha$ and $A, B$ - arbitrary;

(3) $\beta = 16 \alpha$ and $B = 16 A$.

\vspace{2ex}

It seems that Ito \cite{Ito} was the first who rigorously studied the non-integrability of (\ref{hh}) using
the Ziglin's monodromy approach (see  Sect. \ref{theory}). Assuming that $A = B$, he established that there exists an additional
independent integral only in the above cases.
Later, Morales-Ruiz \cite{M} studied the non-integrability using Differential Galois approach. Again assuming $A=B$, he showed that (\ref{hh})
is integrable in Liouvillian sense only in the above cases. Finally, Wenlei Li et al \cite{Wenlei} proved using Differential Galois approach
without assuming priory $A=B$ that (\ref{hh}) is Liouville integrable only in the above cases.

Let us recall some generalizations of (H-H2). In fact, H\'{e}non and Heiles have studied in \cite{HH} an axisymmetric three dimensional natural
Hamiltonian system describing galactic motion. By means of the angular momentum integral they have reduced the considered
system to a two-degrees of freedom one and have carried out numerical experiments in searching for additional integral on the zero level
of the angular momentum integral. Later, Ferrer et al \cite{Fer1,Fer2}, again for three dimensional axisymmetric H\'{e}non-Heiles systems
have studied normal forms, relative equilibria, bifurcations and existence of periodic orbits.

Several multi-dimensional integrable generalizations of the case (2) above are known. In Eilbeck et al \cite{Eilbeck}
a Lax pair for such a generalization is given, and hence, the commuting integrals are obtained. Kostov et al \cite{Kostov} have shown that the complete
integrability is preserved  in \cite{Eilbeck} even considering the additional terms of the kind $\frac{C_j^2}{q_j^2}$ with arbitrary constants $C_j$.
Zeng \cite{Zeng} has constructed a hierarchy of multidimensional H\'{e}non-Heiles systems with the help of the $x$- and $t_n$- constrained flows
of the KdV hierarchy.

In recent paper Fakkousy et al. \cite{Fakkousy} have studied for integrability the following generalization of the H\'{e}non-Heiles system
(H-H3)
\begin{equation}
\label{1.1}
H = \frac{1}{2} (p_1 ^2 + p_2 ^2 + p_3 ^2)
+\frac{1}{2} (A q_1 ^2 + C q_2 ^2 + B q_3 ^2) + (\alpha q_1 ^2 +
\gamma q_2 ^2)q_3 + \frac{\beta}{3}q_3 ^3 ,
\end{equation}
assuming further that $A = C$ and $\alpha = \gamma$. Due to this symmetry there exists an additional integral, this is clear
after introducing spherical coordinates. On the zero level of that integral the Hamiltonian (\ref{1.1}) coincides
with the (H-H2) Hamiltonian (\ref{hh}).
Then, starting from the integrable cases  listed above, the authors have succeeded in finding a complete set
of commuting first integrals for (H-H3) in the following cases:

\vspace{2ex}

(i) $\alpha = \gamma, \quad \frac{\alpha}{\beta} = 1, \quad A = B = C$;

(ii) $\alpha = \gamma, \quad \frac{\alpha}{\beta} = \frac{1}{6}, \quad A = C$, \, $B$-arbitrary;

(iii) $\alpha = \gamma, \quad \frac{\alpha}{\beta} = \frac{1}{16}, \quad A = C, \, \frac{A}{B} = \frac{1}{16}$.

\vspace{2ex}

Again we have to add to the above list the case  $\alpha = \gamma = 0$ in which the Hamiltonian (\ref{1.1}) is separable.
As we point out earlier, the integrability in the second case (ii) can be extended for $A, B, C$ arbitrary.

Numerical study of (H-H3), carried out in \cite{Fakkousy} near the cases of integrability reveals chaotic behavior which
prevents integrability. But there is still a possibility for existence of other integrable cases far from those already found.
The motivation for our study is to establish whether this is true.
Without assuming a priori that  $A = C$ and $\alpha = \gamma$ we establish the following result.

\begin{thm}
\label{th1}
The three dimensional H\'{e}non-Heiles system corresponding to (\ref{1.1}) is non-integrable by means of meromorphic first
integrals except for the cases given above.
\end{thm}

The paper is organized as follows. In Section \ref{theory} we recall some notions and results from what is now called
Ziglin-Morales-Ramis theory, including its generalization with higher order variational equations.
For the reader's convenience, we also add two appendices in which we summarize known results concerning
necessary conditions for integrability of Hamiltonian systems with homogeneous potentials and
of Hamiltonian systems with variational equations of Lam\'e type. In Section \ref{proof}, the proof of Theorem \ref{th1}
is carried out in several steps.

First, we study the necessary conditions for integrability of the homogeneous Hamiltonian related to
(\ref{1.1}). Making use of the results in Appendix A1, we narrow the range of the values of the parameters, for which an
additional integral would exist. In Step 2, already for the initial Hamiltonian (\ref{1.1}),
we study the necessary conditions for integrability of the variational equations along a particular
solution, which are of  Lam\'e type. In this way, we recover the integrability case (iii).
Still, there are values of the parameters for which we can not able to conclude non-integrability. To resolve these cases,
in Step 3 we derive the higher variational equations up to order 3. Then, we use these equations in Steps 4 and 5, and along the way we recover
the integrable case (i). Finally, in Step 6 we present the additional to $H$ commuting integrals in the integrable case
(ii) for $A, B, C$ arbitrary.

Thus, we get a complete answer to the question about integrability and non-integrability of the considered system.

\section{Theory}
\label{theory}

In this section,  we summarize some notions and results related to Ziglin-Morales-Ramis theory. We refer the reader to
\cite{M,MRS1,MR2}) for a more complete exposition on Differential Galois approach to integrability of Hamiltonian systems.
A very detailed presentation about Differential Galois Theory may be found in the book of van der Put and Singer \cite{SvP}).

Given an analytic Hamiltonian $H$, defined on a complex $2n$-dimensional manifold $\mathcal{M}$ determining the system
\begin{equation}
\label{2.1}
\dot{\mathbf{x}} = X_{H} (\mathbf{x}), \quad t \in
\mathbb{C}, \quad \mathbf{x} \in \mathcal{M}
\end{equation}
Recall that such Hamiltonian system is completely integrable (or integrable in Liouville sense) if it admits $n$
independent first integrals $F_1 = H, F_2, \ldots, F_n$ in involution.

Assume that the system (\ref{2.1}) has a non-equilibrium solution $\varphi (t), t \in \hat{I} \subset \mathbb{C}$.
The image of $\hat{I}$ by $\varphi$ is a Riemann surface $\Gamma$. We can write the equation in variation (VE)
along this solution
\begin{equation}
\label{2.2}
\dot{\mathbf{\xi}} = D X_{H} ( \varphi (t))
\mathbf{\xi}, \quad \mathbf{\xi} \in T_{\Gamma} M.
\end{equation}
Further, using the integral $d H$ we can reduce the variational
equation. Consider the normal bundle of $\Gamma$, $\mathcal{F}:= T_{\Gamma}
M / TM$ and let $\pi : T_{\Gamma} M \to \mathcal{F}$ be the natural
projection. The equation (\ref{2.2}) induces an equation on $\mathcal{F}$
\begin{equation}
\label{2.3}
\dot{\mathbf{\eta}} = \pi_{*} (D X_{H} ( \varphi
(t))(\pi^{-1} \mathbf{\eta}) , \quad \mathbf{\eta} \in \mathcal{F}.
\end{equation}
which is  called the normal variational equation (NVE).
Each meromorphic first integral of the Hamiltonian system (\ref{2.1}) in a vicinity
of $\Gamma$ gives rise to a meromorphic first integral of (NVE) \cite{Z1,Z2}.
Therefore, the problem of complete integrability of (\ref{2.1}) reduces to the study of
integrability of the linear systems (\ref{2.2}) or (\ref{2.3}).

Consider such a linear system
\begin{equation}
\label{2.4}
\dot{x} = A (t) x, \quad x \in \mathbb{C}^n, \quad t \in \Gamma .
\end{equation}
The continuation of the solutions along nontrivial loops on $\Gamma$ defines a
linear automorphisms of the space of solutions, called monodromy transformations.
More precisely, let $Y (t)$ be a fundamental solution of (\ref{2.4}), analytic in some neighborhood
of any non-singular point $t_0$. The linear automorphism $\Delta_{\gamma}$, associated with the loop
$\gamma \in \pi_1 (\Gamma, t_0)$ corresponds to a multiplication of $Y (t)$ from the right by a constant matrix
$M_{\gamma}$ - monodromy matrix.
$$
\Delta_{\gamma} Y (t) = Y (t) M_{\gamma}.
$$
The set of these matrices constitute the monodromy group.

In 1982, Ziglin \cite{Z1,Z2} used the relation between the monodromy group of (VE) or (NVE) and branching of solutions
to obtain necessary conditions for integrability of complex Hamiltonian systems. Let us mention one more application of
Ziglin's theory which is related to our study: in 1987, Yoshida \cite{Yoshida} found a criterion for non-existence of
an additional integral in two degrees of freedom Hamiltonian systems with a homogeneous potential.

Morales-Ruiz and Ramis extended the Ziglin's approach in studying the integrability of Hamiltonian systems by investigating
the structure of the differential Galois group of (NVE) ( or (VE)) along certain non-equilibrium solution. We briefly recall some notions
and facts.

Denote the coefficient field in (\ref{2.4}) by $K$.
A differential field $K$ is a field with derivation $\partial =
'$, i.e. an additive mapping satisfying Leibnitz rule. A differential automorphism of $K$ is an
automorphism commuting with the derivation.
Let $y_{i j}$ be the elements of the
fundamental matrix $Y (t)$. Let $L (y_{i j})$ be the extension of
$K$ generated by $K$ and $y_{i j}$ -- a differential field. This
extension is called Picard-Vessiot extension.
 Similarly to classical Galois Theory we define the Galois group
$G := Gal_{K} (L) = Gal (L/K)$ to be the group of all differential
automorphisms of $L$ leaving the elements of $K$ fixed. The Galois
group is, in fact, an algebraic group. It has a unique connected
component $G^0$ which contains the identity and which is a normal
subgroup of finite index. The Galois group $G$ can be represented
as an algebraic linear subgroup of $GL (n, \mathbb{C})$ by
$$
\sigma (Y (t)) = Y (t) R_{\sigma},
$$
where $\sigma \in G$  and $R_{\sigma} \in GL (n, \mathbb{C})$ (see
e.g. \cite{SvP}).

 The solutions of (\ref{2.2}) define an extension $L_1$ of the coefficient field
 $K$ of (VE). This naturally defines a differential Galois group  $G = Gal(L_1/K)$.
  Then, the following fundamental result has been established
 \begin{thm}
 \label{th2}
 {\rm (Morales-Ruiz-Ramis \cite{M})} Suppose that a Hamiltonian system has $n$
  meromorphic first integrals in involution. Then the identity component  $G^0$
  of the Galois group $G = Gal (L_1/K)$ is abelian.
 \end{thm}
If it turns out that $G^0$ is not abelian,  the studied Hamiltonian  system  is non-integrable in the Liouville sense.
But since the above result is only a necessary condition if $G^0$ is abelian  that doesn't imply
integrability.

To obtain other obstacles to the integrability, a method based on
the higher variational equations has been introduced in \cite{M}
and  the previous Theorem has been extended in \cite{MRS1}. Before
formulating this result let us give an idea of higher variational
equations. For the system (\ref{2.2}) with a particular solution
$\varphi (t)$ we put
\begin{equation}
\label{2.5}
\mathbf{x} = \varphi (t) + \varepsilon \xi^{(1)} +
\varepsilon^2 \xi^{(2)} + \ldots + \varepsilon^k \xi^{(k)} +
\ldots,
\end{equation}
where $\varepsilon$ is a formal small parameter. Substituting the
above expression into (\ref{2.1}) and comparing terms with the
same order in $\varepsilon$ we obtain the following chain of
linear non-homogeneous equations
\begin{equation}
\label{2.6}
\dot{\xi}^{(k)} = A (t) \xi^{(k)} + r_k (\xi^{(1)},
\ldots, \xi^{(k-1)}), \quad k = 1, 2, \ldots ,
\end{equation}
where $A (t) = D X_{H} ( \varphi (t))$ and $r_1 \equiv 0$. The
equation (\ref{2.6}) is called k-th variational equation
(${\rm{VE}}_k$). Let $Y (t)$ be the fundamental matrix of
(${\rm{VE}}_1$)
$$
\dot{Y} = A (t) Y.
$$
Then the solutions of $({\rm{VE}}_k), k > 1$ can be found by
\begin{equation}
\label{2.7}
\xi^{(k)} = Y (t) c (t),
\end{equation}
where $c (t)$ is a solution of
\begin{equation}
\label{2.8}
\dot{c} = Y^{-1} (t) r_k .
\end{equation}
As we mention above, the higher variational equations (${\rm{VE}}_k$)
are not  homogeneous equations, but they can be made such,  and therefore, one can define
successive extensions $ K \subset L_1 \subset L_2 \subset \ldots
\subset L_k$, where $L_k$ is the extension obtained by adjoining
the solutions of (${\rm{VE}}_k$). Then, naturally one can define
the Galois groups $Gal (L_1 /K), \ldots, Gal (L_k /K)$. The
following theorem is proven in \cite{MRS1}.
\begin{thm}
\label{th3}
If the Hamiltonian system (\ref{2.1}) is integrable in
Liouville sense then the identity component of every Galois group
$ Gal (L_k /K)$ is abelian.
\end{thm}

Notice that we apply Theorem \ref{th3} in the situation when the
identity component of the Galois group $Gal(L_1/K)$ is abelian.
This means that the first variational equation is solvable. Once
we have the solution of $({\rm{VE}}_1)$, then the solutions of
$({\rm{VE}}_k)$ can be found successively by the method of variations of
constants in the way explained above. In turn, this implies that
the Galois groups $Gal(L_k /K)$ are solvable. One possible way to
show that some of them is not abelian is to find a logarithmic term in the corresponding local
solution (see detailed descriptions and explanations in \cite{M,MRS1,MR2}).

\section{Proof of Theorem \ref{th1}}
\label{proof}

For the proof we use the strategy  from \cite{Wenlei} applied in
studying the non-integrability of the two dimensional H\'{e}non-Heiles system.

Any natural Hamiltonian with a polynomial potential can be represented in the form
\begin{equation}
\label{3.1}
H = \frac{1}{2} \sum p_j ^2 + V (q) = \frac{1}{2} \sum
p_j ^2 + V_{min} (q) + \ldots + V_{max} (q),
\end{equation}
where $V_{min} (q) (V_{max} (q))$ is the lowest (highest) order
term of $V (q)$. It is established by Hietarinta \cite{Hie} and
Maciejevski, Przybylska \cite{MacPrzy} that if the Hamiltonian
system defined by $H$ is integrable, then integrable are its
subsystems
$$
H_{min} = \frac{1}{2} \sum p_j ^2 + V_{min} (q), \qquad H_{max} =
\frac{1}{2} \sum p_j ^2 + V_{max} (q).
$$
On the other hand, if it turns out that some of the above
subsystems is non-integrable, then the original  Hamiltonian
system is non-integrable.

In our case $H_{min}$ is trivially integrable, so we consider
\begin{equation}
\label{3.2}
H = T + V_{max} = \frac{1}{2}(p_1 ^2 + p_2 ^2 + p_3
^2) + (\alpha q_1 ^2 + \gamma q_2 ^2) q_3 + \frac{\beta}{3}q_3 ^3,
\end{equation}
where $\alpha, \beta, \gamma$ are assumed nonzero. After rescaling
$t$ we get
\begin{equation}
\label{3.3}
H  = \frac{1}{2}(p_x ^2 + p_y ^2 + p_z ^2) + (\alpha_1
x^2 + \gamma_1 y^2) z + \frac{1}{3}z^3
\end{equation}
with $\alpha_1 = \alpha/\beta$ and $\gamma_1 = \gamma/\beta$.

First, we study the integrability of the Hamiltonian system
corresponding to the Hamiltonian (\ref{3.3}).

\vspace{2ex}

\noindent
{\bf Step 1. Necessary conditions for the integrability of (\ref{3.3})}

\vspace{2ex}

The equations of motion for the Hamiltonian (\ref{3.3}) are
\begin{equation}
\label{3.4}
\ddot{q}_1  = - 2 \alpha_1 q_1 q_3, \quad \ddot{q}_2 =
- 2 \gamma_1 q_2 q_3, \quad \ddot{q}_3  = - \alpha_1 q_1 ^2 -
\gamma_1 q_2 ^2 - q_3 ^2.
\end{equation}
We are looking for a particular solution of the form
\begin{equation}
\label{3.5}
q_1 = c_1 u (t), \quad  q_2 = c_2 u (t), \quad q_3 = c_3 u (t),
\end{equation}
where
\begin{equation}
\label{3.6}
\ddot{u} = - u^2, \qquad \dot{u}^2 = \frac{2}{3} ( 1 - u^3).
\end{equation}
Then, $\mathbf{c} = (c_1, c_2, c_3)^T$ satisfies the following
system
\begin{equation}
\label{3.7}
c_1  =  2 \alpha_1 c_1 c_3, \quad c_2  =  2 \gamma_1
c_2 c_3, \quad c_3  =  \alpha_1 c_1^2 + \gamma_1 c_2^2 + c_3^2.
\end{equation}
There are several possibilities for solutions of (\ref{3.7}) which
we are going to use:

1) $ c_1 = 0, \quad c_2 = \frac{1}{2 \gamma_1} \sqrt{2 -
\frac{1}{\gamma_1}}, \quad c_3 = \frac{1}{2 \gamma_1}$.

The variational equations (VE) along the found particular solution
read
\begin{equation}
\label{3.8}
\ddot{\mathbf{\xi}} = - u (t) V'' (\mathbf{c})
\mathbf{\xi},
\end{equation}
where $\mathbf{\xi} = (\xi_1, \xi_2, \xi_3)^T$ and
\begin{equation}
\label{3.9}
V'' (\mathbf{c}) =
\begin{pmatrix}
2 \alpha_1 c_3 & 0              & 2 \alpha_1 c_1 \\
0              & 2 \gamma_1 c_3 & 2 \gamma_1 c_2 \\
2 \alpha_1 c_1 & 2 \gamma_1 c_2 & 2 c_3
\end{pmatrix} =
\begin{pmatrix}
\frac{\alpha_1}{\gamma_1} & 0              & 0 \\
0              & 1 &  \sqrt{2 - \frac{1}{\gamma_1}}\\
0 & \sqrt{2 - \frac{1}{\gamma_1}}  & \frac{1}{\gamma_1}
\end{pmatrix}.
\end{equation}
The eigenvalues of $V'' (\mathbf{c})$ are
\begin{equation}
\label{3.10}
\lambda_1 = \frac{\alpha_1}{\gamma_1}, \qquad
\lambda_2 = \frac{1}{\gamma_1} - 1, \qquad \lambda_3 = 2.
\end{equation}

2)  $ c_2 = 0, \quad c_1 = \frac{1}{2 \alpha_1} \sqrt{2 -
\frac{1}{\alpha_1}}, \quad c_3 = \frac{1}{2 \alpha_1}$.

The eigenvalues of $V'' (\mathbf{c})$ in this case are
\begin{equation}
\label{3.11}
\lambda_1 = \frac{\gamma_1}{\alpha_1}, \qquad
\lambda_2 = \frac{1}{\alpha_1} - 1, \qquad \lambda_3 = 2.
\end{equation}

3) $ c_1 = 0, \quad c_2 = 0, \quad c_3 = 1$.

The eigenvalues of $V'' (\mathbf{c})$ in this case are
\begin{equation}
\label{3.12}
\lambda_1 = 2 \alpha_1, \qquad \lambda_2 = 2
\gamma_1, \qquad \lambda_3 = 2.
\end{equation}

Now,  if the Hamiltonian system is integrable, the pairs $(k=3,
\lambda_j)$ have to belong to one of the following cases (1),
(11), (12), (13), (14), (18) of Theorem \ref{thA1} (see Appendix A1).

Denote
$$g_1 (s) = s + s(s-1)3/2, \quad g_2 (s) = -\frac{1}{24} + \frac{1}{24} ( 2 + 6 s )^2, \quad
g_3 (s) = -\frac{1}{24} + \frac{1}{24} \left( \frac{3}{2} + 6 s
\right)^2 ,$$
$$ g_4 (s) = -\frac{1}{24} + \frac{1}{24} \left( \frac{6}{5} + 6 s \right)^2, \quad
g_5 (s) = -\frac{1}{24} + \frac{1}{24} \left( \frac{12}{5} + 6 s
\right)^2 , \quad g_6 (s) = \frac{1}{3} + \frac{3}{2} s (s+1)$$
 and observe that
$g _j (s) > 0$ for $s \in \mathbb{Z}$. Then, if $\lambda_2 = g_j
(s)$ for some integer $s$ in (\ref{3.10}), (\ref{3.11}) we get
that
\begin{equation}
\label{3.13}
\alpha_1, \gamma_1 \in (0, 1], \qquad \alpha_1,
\gamma_1 \in \mathbb{Q}.
\end{equation}

Now, to obtain necessary conditions for existence of additional
first integral we incorporate the condition (\ref{s9}) from
Theorem \ref{thA2} (see Appendix A1) for $\lambda_1, \lambda_2$ in any of above
possibilities. We have
\begin{equation}
\label{3.14}
\frac{1}{6} \sqrt{1 + 24 \frac{\alpha_1}{\gamma_1} }
= \frac{1}{6} \sqrt{1 + 24 \left(\frac{1}{\gamma_1} -1\right) } +
l_1,
\end{equation}
\begin{equation}
\label{3.15}
\frac{1}{6} \sqrt{1 + 24 \frac{\gamma_1}{\alpha_1} }
= \frac{1}{6} \sqrt{1 + 24 \left(\frac{1}{\alpha_1} -1\right) } +
l_2,
\end{equation}
\begin{equation}
\label{3.16}
\frac{1}{6} \sqrt{1 + 48\alpha_1} = \frac{1}{6}
\sqrt{1 + 48\gamma_1}  + l_3,
\end{equation}
for some $l_1, l_2, l_3 \in \mathbb{Z}$. It is known from
\cite{MPY} that all radicals in these expressions are in fact
rational numbers.

Consider the relation (\ref{3.16}). Due to (\ref{3.13})
$$
1 < \sqrt{1 + 48\alpha_1}, \,  \frac{1}{6} \sqrt{1 + 48\gamma_1}
\leq 7.
$$
Hence, (\ref{3.16}) is valid only for $l_3 = 0$, but this gives
$$
\gamma_1 = \alpha_1.
$$

Alternatively, the equality  $\gamma_1 = \alpha_1$ can be obtained
also by exploring the necessary conditions from Theorem \ref{thA1}.
Suppose $\lambda_1 = \frac{\alpha_1}{\gamma_1} := w$ from
(\ref{3.10}) takes values in some $g_i (s), i=1, \ldots, 6, s \in
\mathbb{Z}$, i.e., $w = g_i (s)$. Similarly, for integrability
$\lambda_1 = \frac{\gamma_1}{\alpha_1} :=\frac{1}{w}$ from
(\ref{3.11}) also has to take values in some $g_j (q), j=1,
\ldots, 6, q \in \mathbb{Z}$, that is, $\frac{1}{w} = g_j (q)$.
Then we must have
$$
g_j (q) = \frac{1}{g_i (s)}
$$
for some $i, j=1, \ldots, 6$ and some $q, s \in \mathbb{Z}$. But
this can be achieved only for $i = j = 1$ and $q = s = 1$, which
implies that $w=1$, or $\gamma_1 = \alpha_1$.

This result has two major consequences: firstly, the Hamiltonian
system (\ref{3.4}) admits an additional integral $F = q_1 p_2 -
q_2 p_1$, and secondly, we get a representation from (\ref{3.15})
for  $\alpha_1$ for which we may have complete integrability,
namely
\begin{equation}
\label{3.17}
\frac{5}{6} = \frac{1}{6} \sqrt{1 + 24
\left(\frac{1}{\alpha_1}-1\right) } + l, \quad l \in \mathbb{Z},
\end{equation}
which is equivalent to
\begin{equation}
\label{3.18}
\frac{1}{\alpha_1}-1 = - \frac{1}{24} + \frac{1}{24}
(5 - 6 l)^2 \quad \mbox{or} \quad \alpha_1 = \frac{24}{23 + (5-6
l)^2}.
\end{equation}

We will use the above representation to obtain those $\alpha_1$
for which the Hamiltonian system (\ref{3.4}) is necessarily
integrable by investigating whether there are  some $s, l \in
\mathbb{Z}$, such that $\lambda_2$ from (\ref{3.11}) and
(\ref{3.12}) takes values in the six families $g _j (s), j = 1, 2,
\ldots, 6$. Starting with $\lambda_2$ from (\ref{3.11}) we have
\begin{equation}
\label{3.19} \frac{1}{\alpha_1}-1 = - \frac{1}{24} + \frac{1}{24}
(5 - 6 l)^2 = g_j (s), \quad j = 1, \ldots, 6.
\end{equation}
Trivial computations show that there are no such integers $s, l$
for $j = 2, \ldots, 6$. However,
$$
- \frac{1}{24} + \frac{1}{24} (5 - 6 l)^2 = g_1 (s) = \frac{3}{2}
s^2 - \frac{1}{2} s
$$
is reduced to $l + s = 1$, which in turn gives
\begin{itemize}
\item $s = 0, l = 1$, and therefore, $\alpha_1 = 1$; \item $s = 1,
l = 0$, and therefore, $\alpha_1 = \frac{1}{2}$.
\end{itemize}

Further, we turn to $\lambda_2$ from (\ref{3.12}). Using that
$\alpha_1 = \gamma_1$ and (\ref{3.18}), we have to find $s, l \in
\mathbb{Z}$  for which the following relation
\begin{equation}
\label{3.20}
 2 \alpha_1 = \frac{48}{23 + (5 -6 l)^2} = g_j (s), \quad \mbox{for some} \, \, s , l \in \mathbb{Z} .
\end{equation}
is fulfilled. Since $0 <\frac{48}{23 + (5 -6 l)^2} \leq 1$, we
should have $0 < g_j (s) \leq 1$ and this gives a very few
possibilities for $s$. Indeed,
$$
\frac{48}{23 + (5 -6 l)^2} = g_1 (s) = \frac{3}{2} s^2 -
\frac{1}{2} s
$$
is fulfilled for $s=0$ and $l=1$, which amounts to $\alpha_1 =
1/2$, but we know that.

Next,
$$
\frac{48}{23 + (5 -6 l)^2} = g_2 (s) = -\frac{1}{24}  +
\frac{1}{24} (6 s+2)^2
$$
is fulfilled for $s=0$ and $l=4$, which gives $\alpha_1 =
\frac{1}{16}$. There are no $s, l \in \mathbb{Z}$ which satisfy
$$
\frac{48}{23 + (5 -6 l)^2} = g_j (s), \quad j = 3, 4, 5.
$$
Finally,
$$
\frac{48}{23 + (5 -6 l)^2} = g_6 (s) = \frac{1}{2} \Big[
\frac{2}{3} + 3 s (s+1) \Big]
$$
is fulfilled for $s=0$ and $l=-1$, and hence, $\alpha_1 =
\frac{1}{6}$.

Summarizing, the necessary conditions for the system (\ref{3.4})
to be integrable are $\alpha_1 = \gamma_1$ and $\alpha_1 = 1,
\frac{1}{2}, \frac{1}{6}, \frac{1}{16}$. Therefore, we have the
following

\begin{prop} The Hamiltonian system (\ref{3.4}) corresponding to the Hamiltonian with the homogeneous potential
(\ref{3.3}) is non-integrable if

(i) $\gamma_1 \neq \alpha_1$, or

(ii) $\gamma_1 = \alpha_1, \quad \alpha_1 \in \mathbb{C} \setminus
\left\{1, \frac{1}{2},  \frac{1}{6},  \frac{1}{16} \right\}$

\end{prop}

$\hfill \square$

\vspace{3ex}

Now, consider the original Hamiltonian (\ref{1.1}) assuming that
$A, B$ and $C$ are nonzero. Again after rescaling $t$ we get
\begin{equation}
\label{3.30}
H = \frac{1}{2} (p_1 ^2 + p_2 ^2 + p_3 ^2)
+\frac{1}{2} (A_1 q_1 ^2 + C_1 q_2 ^2 + B_1 q_3^2) + (\alpha_1 q_1
^2 + \gamma_1 q_2 ^2)q_3 + \frac{1}{3} q_3 ^3,
\end{equation}
where $A_1 = A/\beta, B_1 = B/\beta, C_1 = C/\beta$ and $\alpha_1
= \gamma_1$ are as before.
The Hamiltonian system  corresponding to (\ref{3.30}) reads
\begin{equation}
\label{Hequ}
\ddot{q}_1 = -A_1 q_1 - 2 \alpha_1 q_1 q_3, \quad \ddot{q}_2 = -C_1 q_2 - 2 \alpha_1 q_2 q_3, \quad
\ddot{q}_3 = -B_1 q_3 - \alpha_1 (q_1^2 + q_2^2) - q_3^2.
\end{equation}
The above system admits the following particular solution
\begin{equation}
\label{3.31}
\Gamma_h : q_1  = q_2 = p_1 = p_2 = 0, \quad q_3  =
q_3 ^0 = -\frac{1}{2} B_1 + \wp (\tau, g_2, g_3), \quad p_3 =
\dot{q}_3 ^0 ,
\end{equation}
where $\tau = \frac{i}{\sqrt{6}} t, \, i^2 = -1, \, \,  g_2 = 3
B_1 ^2, \, \, g_3 =  12 h - B_1 ^3$.

The variational equation (VE) along this solution, written with
respect to the independent variable $\tau$, ($' = d/d \tau$) are
\begin{eqnarray}
\label{3.32}
\xi_1 '' & - & [12 \alpha_1 \wp (\tau) + 6(A_1 - B_1 \alpha_1)] \xi_1 = 0, \\
\label{3.33}
\xi_2 '' & - & [12 \alpha_1 \wp (\tau) + 6(C_1 - B_1 \alpha_1)] \xi_2 = 0, \\
\label{3.34}
\xi_3 '' & - & 12  \wp (\tau)  \xi_3 = 0.
\end{eqnarray}
Clearly, the equation (\ref{3.34}) forms the tangential part of
(VE), while the equations (\ref{3.32}) and (\ref{3.33}) form
(NVE). Notice that (NVE) are given by two independent second order
equations of Lam\'{e} type.

\vspace{2ex}

\noindent
{\bf Step 2. Analysis of (NVE)}

\vspace{2ex}

In what follows, we apply the results described in the Appendix
A.2. Consider first (\ref{3.32}) written as
$$
\xi_1 '' - f_1 (\tau, h) \xi_1 = 0
$$
with $ f_1 (\tau, h) = 12 \alpha_1 \wp (\tau) + 6(A_1 - B_1
\alpha_1). $ After some calculations we get
\begin{align}
\label{3.35}
f_1 '^2 = P (f_1, h) & = \frac{1}{3 \alpha_1} f_1^3 + \frac{6}{\alpha_1}(A_1 - B_1 \alpha_1) f_1^2 + [-36 \alpha_1 B_1^2 + \frac{36}{\alpha_1}(A_1 - B_1 \alpha_1)^2]f_1 \\
                  & - 216 \alpha_1 B_1^2 (A_1 - B_1 \alpha_1) + \frac{72}{\alpha_1}(A_1 - B_1 \alpha_1)^3 + (12)^2 \alpha_1^2 B_1^3 - h (12)^3 \alpha_1^2. \nonumber
\end{align}
We write down the coefficients of the polynomial $P (f_1, h)$ in a way we need them
\begin{align*}
a_1 & = \frac{1}{3 \alpha_1}, & a_2 = 0, \\
b_1 & = \frac{6}{\alpha_1} (A_1 + B_1 \alpha_1), & b_2 = 0, \\
c_1 & = -36 \alpha_1 B_1^2 + \frac{36}{\alpha_1} (A_1 - B_1 \alpha_1)^2, & c_2 = 0, \\
d_1 & = - 216 \alpha_1 B_1^2 (A_1 - B_1 \alpha_1)+
\frac{72}{\alpha_1}(A_1 - B_1 \alpha_1)^3 + (12)^2 \alpha_1^2
B_1^3, & d_2 = -(12)^3 \alpha_1^2.
\end{align*}

Now, we explore the necessary conditions for integrability as they
are given in the Appen\-dix A2. The first necessary condition from
Theorem \ref{thA2} -- ${\rm I}. \, a_1 = \frac{4}{n(n+1)}$ for
some $n \in \mathbb{N}$ gives that $\alpha_1 = \frac{n(n+1)}{12}$
for some $n \in \mathbb{N}$. Recall that we have to deal only with
the cases
\begin{equation}
\label{3.37}
n = 1 \, \, \left(\alpha_1 = \frac{1}{6}\right),
\quad n = 2 \, \, \left(\alpha_1 = \frac{1}{2}\right), \quad
\mbox{and} \quad n = 3 \, \, (\alpha_1 = 1)
\end{equation}
because of Proposition 1.

Next, we proceed with the cases of condition ${\rm II}.$

The case ${\rm II}_1.$ $m = 1$ gives that $\alpha_1 =
\frac{1}{16}$, while the condition $b_1 = 0$ amounts to
\begin{equation}
\label{3.38}
\frac{A_1}{B_1} =  \alpha_1 = \frac{1}{16} \,
\left(\frac{A}{B} = \frac{1}{16} \right).
\end{equation}
The case ${\rm II}_2.$ $m = 2$ gives $\alpha_1 = 5/16$, but for
values of $\alpha_1$ different from $\{1, 1/2, 1/6, 1/16 \} $ we
have non-integrability again due to the results from Step 1.

The case ${\rm II}_3.$ $m = 3$ does not occur here since $c_2 = 0$
and $16 a_{1} d_2 + 3 b_{1} c_2 = 0$ yield $a_{1} d_{2} = 0$,
which is not possible.

The case ${\rm II}_m.$ $m > 3$ is also not possible due to
$d_2 \neq 0$ and the assumptions $A_1 \neq 0$ and $B_1 \neq 0$.

Finally, the case  ${\rm III}$ does not occur, because after
expressing all necessary conditions we get $B_1 = 0$ which is a
contradiction to our assumption.

Similarly, (\ref{3.33}) can be written as
$$
\xi_2 '' - f_2 (\tau, h) \xi_1 = 0
$$
with $ f_2 (\tau, h) = 12 \alpha_1 \wp (\tau) + 6(C_1 - B_1
\alpha_1)$. The polynomial $P (f_2, h)$ is exactly as (\ref{3.35})
with $A_1$ replaced by $C_1$. The necessary conditions ${\rm I,
II_2, II_3, II_m}$ and ${\rm III}$ give nothing new. In the case
${\rm II_1}$ where $m=1$ and $\alpha_1 = 1/16$ we obtain from the
condition $b_1 = 0$ that
\begin{equation}
\label{3.39}
\frac{C_1}{B_1} =  \alpha_1 = \frac{1}{16} \quad
\left(\frac{C}{B} = \frac{1}{16} \right).
\end{equation}
This condition, together with (\ref{3.38}) results in
\begin{equation}
\label{3.40}
A = C, \qquad \, \frac{A}{B} = \frac{1}{16}, \quad \alpha = \gamma, \quad \frac{\alpha}{\beta} = \frac{1}{16},
\end{equation}
but we know that in this case the Hamiltonian (\ref{1.1}) is integrable.

Summarizing, we get a more definitive answer for the case $\alpha_1 = 1/16$: together with already established condition
$\alpha_1 = \gamma_1 (\alpha = \gamma)$ the Hamiltonian (\ref{1.1}) is necessarily integrable if $A_1 = C_1 (A = C)$,
but this gives the already known integrable case (iii) from the Introduction.

\vspace{2ex}

\noindent
{\bf Step 3. Higher variational equations}

\vspace{2ex}

Further, it remains to deal with the cases (\ref{3.37}). In any of these cases the Galois group of variational equation is abelian
(see Apppendix A.2). Hence,  we need to study the Galois groups of higher variational equations.
As it was explained in Section 2,  to  write them we put
\begin{eqnarray}
\label{ext}
q_1 &=& 0 + \varepsilon \xi_1 ^{(1)} + \varepsilon^2 \xi_1 ^{(2)} +  \varepsilon^3 \xi_1 ^{(3)} + \ldots , \\ \nonumber
q_2 &=& 0 + \varepsilon \xi_2 ^{(1)} + \varepsilon^2 \xi_2 ^{(2)} +  \varepsilon^3 \xi_2 ^{(3)} + \ldots , \\
q_3 &=& q_3^0  + \varepsilon \xi_3 ^{(1)} + \varepsilon^2 \xi_3 ^{(2)} +  \varepsilon^3 \xi_3 ^{(3)} + \ldots  \nonumber
\end{eqnarray}
and substitute these expressions into (\ref{Hequ}). Comparing the terms with the same order in $\varepsilon$, we obtain the variational
equations up to order three.

We are going to write these higher variational equations as second-order linear equations with respect to the independent variable
$\tau$. The first variational equation (${\rm VE}_1$) is nothing but (\ref{3.32}) -- (\ref{3.34})
\begin{eqnarray}
\label{3.41}
(\xi_1 ^{(1)}) '' & - & [12 \alpha_1 \wp (\tau) + 6(A_1 - B_1 \alpha_1)] \xi_1 ^{(1)} = 0, \\
\label{3.42}
(\xi_2 ^{(1)}) '' & - & [12 \alpha_1 \wp (\tau) + 6(C_1 - B_1 \alpha_1)] \xi_2 ^{(1)} = 0, \\
\label{3.43}
(\xi_3 ^{(1)}) '' & - & 12  \wp (\tau)  \xi_3 ^{(1)} = 0.
\end{eqnarray}
For the second variational equation (${\rm VE}_2$) we get
\begin{eqnarray}
\label{3.44}
(\xi_1 ^{(2)}) '' & - & [12 \alpha_1 \wp (\tau) + 6(A_1 - B_1 \alpha_1)] \xi_1 ^{(2)} = W_1 ^{(2)}, \\
\label{3.45}
(\xi_2 ^{(2)}) '' & - & [12 \alpha_1 \wp (\tau) + 6(C_1 - B_1 \alpha_1)] \xi_2 ^{(2)} = W_2 ^{(2)}, \\
\label{3.46}
(\xi_3 ^{(2)}) '' & - & 12  \wp (\tau)  \xi_3 ^{(2)} = W_3 ^{(2)},
\end{eqnarray}
where
\begin{equation}
\label{3.47}
W_j ^{(2)} = 12 \alpha_1 \xi_j ^{(1)} \xi_3 ^{(1)}, j = 1,2, \quad W_3 ^{(2)} = 6 \alpha_1 [(\xi_1 ^{(1)})^2 + (\xi_2 ^{(1)})^2 ] + 6 (\xi_3 ^{(1)})^2.
\end{equation}
The third variational equation (${\rm VE}_3$) is
\begin{eqnarray}
\label{3.48}
(\xi_1 ^{(3)}) '' & - & [12 \alpha_1 \wp (\tau) + 6(A_1 - B_1 \alpha_1)] \xi_1 ^{(3)} = W_1 ^{(3)}, \\
\label{3.49}
(\xi_2 ^{(3)}) '' & - & [12 \alpha_1 \wp (\tau) + 6(C_1 - B_1 \alpha_1)] \xi_2 ^{(3)} = W_2 ^{(3)}, \\
\label{3.50}
(\xi_3 ^{(3)}) '' & - & 12  \wp (\tau)  \xi_3 ^{(3)} = W_3 ^{(3)}.
\end{eqnarray}
Here
\begin{equation}
\label{3.51}
W_j ^{(3)} = 12 \alpha_1 (\xi_j ^{(1)} \xi_3 ^{(2)} + \xi_j ^{(2)} \xi_3 ^{(1)}), j = 1,2.
\end{equation}
We do not need $W_3 ^{(3)}$, but it is calculated to be
$W_3 ^{(3)} = 12 \alpha_1 (\xi_1 ^{(1)} \xi_1 ^{(2)} + \xi_2 ^{(1)} \xi_2 ^{(2)}) + 12 \xi_3 ^{(1)} \xi_3 ^{(2)}$.
Keeping our notation from Section 2 we have
\begin{equation}
\label{3.52}
r_2 = (0, W_1 ^{(2)}, 0, W_2 ^{(2)}, 0, W_3 ^{(2)}), \quad  r_3 = (0, W_1 ^{(3)}, 0, W_2 ^{(3)}, 0, W_3 ^{(3)}).
\end{equation}

To prove non-integrability of (\ref{3.30}) we have to show that the identity component of the Galois group of
(${\rm VE}_k, k = 2, 3$) is not abelian. To get this let us note that  the variational equations have a singular point at
$\tau = 0$. Then it is enough to obtain a logarithm around the singular point,  or equivalently, a residue different from zero in some integrand
(see  \cite{M}).
Now we are going to calculate the local solutions around $\tau = 0$ and to show that a logarithmic term appears in the solutions of (${\rm VE}_3$).
Notice that all expansions below are convergent in some vicinity of $\tau = 0$ \cite{WW}.

Let $\xi_{j,1}^{(1)}, \xi_{j,1}^{(1)}, j = 1, 2, 3$ be two linearly independent solutions of (\ref{3.41})-(\ref{3.43}) with unit Wronskian.
Then, the fundamental matrix $Y (\tau)$ of (${\rm VE}_1$) and its inverse have the block-diagonal form
\begin{equation}
\label{3.53}
Y (\tau) = \begin{pmatrix}
Y_1 & 0   & 0 \\
0   & Y_2 & 0 \\
0   & 0   & Y_3
\end{pmatrix},
\qquad
Y^{-1} (\tau) = \begin{pmatrix}
Y_1 ^{-1} & 0 & 0 \\
0 & Y_2 ^{-1} & 0 \\
0 & 0 &  Y_3 ^{-1}
\end{pmatrix},
\end{equation}
where
\begin{equation}
\label{3.54}
Y_j = \begin{pmatrix}
\xi_{j,1} ^{(1)}     & \xi_{j,2} ^{(1)} \\
(\xi_{j,1} ^{(1)})'  & (\xi_{j,2} ^{(1)})'
\end{pmatrix},
\qquad
Y_j ^{-1} = \begin{pmatrix}
(\xi_{j,2} ^{(1)})' & -\xi_{j,2} ^{(1)} \\
-(\xi_{j,1} ^{(1)})' & \xi_{j,1} ^{(1)}
\end{pmatrix}.
\end{equation}

Now, we return to the integrability analysis of the cases (\ref{3.37}).

\vspace{2ex}

\noindent
{\bf Step 4. Non-integrability of the case $n = 3, \quad \alpha_1 = 1$.}

\vspace{2ex}

One can find the following expansions of the solutions of (${\rm VE}_1$) (\ref{3.41}), (\ref{3.42}) and (\ref{3.43}) with $\alpha_1 = 1$.
\begin{eqnarray}
\label{4.1}
\xi_{1,1} ^{(1)} & = & \frac{1}{\tau^3} \Big[1 - \frac{3}{5}\left(A_1 -B_1\right)\tau^2 + \frac{3}{10}\left(A_1 ^2 - 2 A_1 B_1 + \frac{B_1^2}{2}\right) \tau^4 + \ldots \Big], \nonumber \\
\xi_{1,2} ^{(1)} & = & \frac{\tau^4}{7} \Big[1 + \frac{1}{3} (A_1 - B_1) \tau^2 + \ldots \Big].
\end{eqnarray}
Similarly, we have
\begin{eqnarray}
\label{4.2}
\xi_{2,1} ^{(1)} & = & \frac{1}{\tau^3} \Big[1 - \frac{3}{5}\left(C_1 -B_1\right)\tau^2 + \frac{3}{10}\left(C_1 ^2 - 2 C_1 B_1 + \frac{B_1^2}{2}\right) \tau^4 + \ldots \Big], \nonumber \\
\xi_{2,2} ^{(1)} & = & \frac{\tau^4}{7} \Big[1 + \frac{1}{3} (C_1 - B_1) \tau^2 + \ldots \Big]
\end{eqnarray}
and also,
\begin{equation}
\label{4.3}
\xi_{3,1} ^{(1)}  =  \frac{1}{\tau^3} - \frac{g_2}{20} \tau - \frac{g_3}{14} \tau^3 + \frac{g_2 ^2}{400} \tau^5 + \ldots, \qquad
\xi_{3,2} ^{(1)}  =  \frac{\tau^4}{7} \Big[1 + \frac{3}{220} g_2 \tau^4 + \ldots \Big] .
\end{equation}

There are no logarithmic terms in the expansions around $\tau = 0$ of the local solutions of (${\rm VE}_2$) and
in what follows we write down the expansions of their local solutions  only in the cases we will use further.

For the first equation of (${\rm VE}_2$) with a specific choice of the right-hand terms
\begin{equation}
\label{4.4}
(\xi_1 ^{(2)}) ''  -  [12 \alpha_1 \wp (\tau) + 6(A_1 - B_1)] \xi_1 ^{(2)} = 12 \xi_{1,2} ^{(1)} \xi_{3,1} ^{(1)}
\end{equation}
we find the following expansions around $\tau=0$
\begin{equation}
\label{4.5}
\xi_{1,j} ^{(2)}  =  \xi_{1,j} ^{(1)} + \tau^3 \Big[- \frac{2}{7} - \frac{1}{7}(A_1-B_1) \tau^2 + \ldots \Big], \quad j=1,2.
\end{equation}
Similarly, for the second equation
\begin{equation}
\label{4.6}
(\xi_2 ^{(2)}) ''  -  [12 \alpha_1 \wp (\tau) + 6(C_1 - B_1)] \xi_2 ^{(2)} = 12 \xi_{2,2} ^{(1)} \xi_{3,1} ^{(1)}
\end{equation}
we obtain
\begin{equation}
\label{4.7}
\xi_{2,j} ^{(2)}  =  \xi_{2,j} ^{(1)} + \tau^3 \Big[- \frac{2}{7} - \frac{1}{7}(C_1-B_1) \tau^2 + \ldots \Big], \quad j = 1,2 .
\end{equation}
Finally, for the third equation in (${\rm VE}_2$) with the following choice of the right-hand terms
\begin{equation}
\label{4.8}
(\xi_3 ^{(2)}) ''  -  12  \wp (\tau)  \xi_3 ^{(2)} = 6 \Big[ (\xi_{1,1} ^{(1)})^2 + (\xi_{2,1} ^{(1)})^2 + (\xi_{3,1} ^{(1)})^2 \Big]
\end{equation}
we get the expansions
\begin{equation}
\label{4.9}
\xi_{3,j} ^{(2)}  =  \xi_{3,j} ^{(1)} + \frac{1}{\tau^4} \Big[\frac{9}{4} + \frac{6}{5}(C_1+A_1 - 2B_1) \tau^2 + K_4 \tau^4 + \ldots \Big], \quad j = 1,2,
\end{equation}
where $K_4$ is
$$
K_4 := -\frac{339}{400}B_1^2 - \frac{12}{25}C_1^2 + \frac{24}{25} A_1 B_1 + \frac{24}{25} C_1 B_1 -\frac{12}{25} A_1^2 .
$$

Now we will study the local solutions of (${\rm VE}_3$).
Consider the first equation with the following specific representatives on the right-hand side.
\begin{equation}
\label{4.10}
(\xi_1 ^{(3)}) ''  -  [12 \alpha_1 \wp (\tau) + 6(A_1 - B_1)] \xi_1 ^{(3)} = 12 \left(\xi_{1,1} ^{(1)} \xi_{3,2} ^{(2)} + \xi_{1,2} ^{(2)} \xi_{3,2} ^{(1)}\right) = W_1 ^{(3)}.
\end{equation}
Denote
$$(\nu_{1,1} ^{(3)}, \nu_{1,2} ^{(3)}) = Y_1 ^{-1}
\begin{pmatrix}
0 \\
W_1 ^{(3)}
\end{pmatrix} .
$$
Then, for $\nu_{1,1} ^{(3)} = -12 \xi_{1,2} ^{(1)} \Big[ \xi_{1,1} ^{(1)} \xi_{3,2} ^{(2)} + \xi_{1,2} ^{(2)} \xi_{3,2} ^{(1)} \Big]$ we obtain
\begin{equation}
\label{4.11}
{\rm Res}_{\tau=0} \, \nu_{1,1} ^{(3)} = - \frac{36}{35} (A_1 + 2 C_1 -  B_1).
\end{equation}
Similar computations for (\ref{3.49})
\begin{equation}
\label{4.12}
(\xi_2 ^{(3)}) ''  -  [12 \alpha_1 \wp (\tau) + 6(C_1 - B_1)] \xi_2 ^{(3)} = 12 \left(\xi_{2,1} ^{(1)} \xi_{3,2} ^{(2)} + \xi_{12,2} ^{(2)} \xi_{3,2} ^{(1)}\right) = W_2 ^{(3)}
\end{equation}
give
$$(\nu_{2,1} ^{(3)}, \nu_{2,2} ^{(3)}) = Y_2 ^{-1}
\begin{pmatrix}
0 \\
W_2 ^{(3)}
\end{pmatrix}
$$
and
\begin{equation}
\label{4.13}
{\rm Res}_{\tau=0} \, \nu_{2,1} ^{(3)} = - \frac{36}{35} (C_1 + 2 A_1 -  B_1).
\end{equation}
Suppose for a moment that the both residues (\ref{4.11}) and (\ref{4.13}) are zero, that is,
$$
A_1 + 2 C_1 - 3 B_1 = 0, \quad C_1 + 2 A_1 - 3 B_1 = 0 .
$$
The only solution of this system is $A_1 = B_1 = C_1 (A = B = C)$, but for these values of the parameters we recover
the already known integrable case (i) from the Introduction.

On the other hand, if $A_1 \neq B_1 \neq C_1$, then at least one of the above residues is non-trivial.
Thus, we have obtained a nonzero residue at $\tau = 0$, which implies a logarithm in the solutions of (${\rm VE}_3$). Then, the
Galois group of (${\rm VE}_3$) is solvable, but not abelian. Hence, the non-integrability of the Hamiltonian system (\ref{Hequ})
in this case follows from Theorem \ref{th3}.

\vspace{2ex}

\noindent
{\bf Step 5. Non-integrability of the case $n = 2, \quad \alpha_1 = \frac{1}{2}$.}

\vspace{2ex}

Here we will show that for $\alpha_1 = 1/2$ the Hamiltonian system (\ref{Hequ}) is not integrable,
whatever the other parameters are. The same line of considerations as in the previous Step is taken.

One can find the following expansions of the solutions of (${\rm VE}_1$) (\ref{3.41}), (\ref{3.42}) and (\ref{3.43}) with $\alpha_1 = \frac{1}{2}$.
\begin{eqnarray}
\label{5.1}
\xi_{1,1} ^{(1)} & = & \frac{1}{\tau^2} \Big[1 - \left(A_1 - \frac{B_1}{2}\right)\tau^2 + \frac{3}{2}\left(A_1 ^2 -  A_1 B_1 + \frac{B_1^2}{10}\right) \tau^4 + \ldots \Big], \nonumber \\
\xi_{1,2} ^{(1)} & = & \frac{\tau^3}{5} \Big[1 + \frac{3}{7} \left(A_1 - \frac{B_1}{2}\right) \tau^2 + \ldots \Big].
\end{eqnarray}
To get the expansions of the solutions of (\ref{3.42}) we only need to replace $A_1$ with $C_1$.
\begin{eqnarray}
\label{5.2}
\xi_{2,1} ^{(1)} & = & \frac{1}{\tau^2} \Big[1 - \left(C_1 - \frac{B_1}{2}\right)\tau^2 + \frac{3}{2}\left(C_1 ^2 -  C_1 B_1 + \frac{B_1^2}{10}\right) \tau^4 + \ldots \Big], \nonumber \\
\xi_{2,2} ^{(1)} & = & \frac{\tau^3}{5} \Big[1 + \frac{3}{7} \left(C_1 - \frac{B_1}{2}\right) \tau^2 + \ldots \Big].
\end{eqnarray}
Notice that we already have the expansion for $\xi_{3} ^{(1)}$ in (\ref{4.3}).

There are no logarithmic terms in the expansions around $\tau = 0$ of the local solutions of (${\rm VE}_2$).
We calculate the expansions of the solutions of (${\rm VE}_2$) only in the cases we need in the sequel.

For the first equation of (${\rm VE}_2$) with a specific choice of the right-hand terms
\begin{equation}
\label{5.3}
(\xi_1 ^{(2)}) ''  -  \Big[6 \alpha_1 \wp (\tau) + 6 \left(A_1 - \frac{B_1}{2}\right)\Big] \xi_1 ^{(2)} = 6 \xi_{1,1} ^{(1)} \xi_{3,2} ^{(1)}
\end{equation}
we find the following expansions
\begin{equation}
\label{5.4}
\xi_{1,j} ^{(2)}  =  \xi_{1,j} ^{(1)} + \tau^4 \Big[\frac{1}{7} + \frac{1}{14}(A_1-B_1/2) \tau^2 + \ldots \Big], \quad j = 1, 2.
\end{equation}
Similarly, for the second equation
\begin{equation}
\label{5.5}
(\xi_2 ^{(2)}) ''  -  \Big[6 \alpha_1 \wp (\tau) + 6 \left(C_1 - \frac{B_1}{2}\right)\Big] \xi_2 ^{(2)} = 6 \xi_{2,1} ^{(1)} \xi_{3,2} ^{(1)}
\end{equation}
we obtain
\begin{equation}
\label{5.6}
\xi_{2,j} ^{(2)}  =  \xi_{1,j} ^{(1)} + \tau^4 \Big[\frac{1}{7} + \frac{1}{14}(C_1-B_1/2) \tau^2 + \ldots \Big], \quad j = 1, 2.
\end{equation}
Finally, for the third equation in (${\rm VE}_2$) with the following choice of the right-hand terms
\begin{equation}
\label{5.7}
(\xi_3 ^{(2)}) ''  -  12  \wp (\tau)  \xi_3 ^{(2)} = 3 \Big[ (\xi_{1,1} ^{(1)})^2 + (\xi_{2,1} ^{(1)})^2 \Big] + 6 (\xi_{3,2} ^{(1)})^2
\end{equation}
we get the expansions
\begin{equation}
\label{5.8}
\xi_{3,j} ^{(2)} =  \xi_{3,j} ^{(1)} + \frac{1}{\tau^2} \Big[-1 + \frac{1}{2}(A_1 - B_1 + C_1) \tau^2  + \ldots \Big], \quad j=1,2.
\end{equation}

It is enough for our purposes to consider only the first equation of (${\rm VE}_3$)
with the following choice of the right-hand terms.
\begin{equation}
\label{5.9}
(\xi_1 ^{(3)}) ''  -  \Big[6 \alpha_1 \wp (\tau) + 6 \left(A_1 - \frac{B_1}{2}\right)\Big] \xi_1 ^{(3)} =
6 \Big[ \xi_{1,2} ^{(1)} \xi_{3,2} ^{(2)} + \xi_{1,2} ^{(2)} \xi_{3,1} ^{(1)} \Big] = W_1 ^{(3)}.
\end{equation}
Denoting again
$$(\nu_{1,1} ^{(3)}, \nu_{1,2} ^{(3)}) = Y_1 ^{-1}
\begin{pmatrix}
0 \\
W_1 ^{(3)}
\end{pmatrix}
$$
we have, this time for $\nu_{1,2} ^{(3)} = 6 \xi_{1,1} ^{(1)} \Big[ \xi_{1,2} ^{(1)} \xi_{3,2} ^{(2)} + \xi_{1,2} ^{(2)} \xi_{3,1} ^{(1)} \Big]$
\begin{equation}
\label{5.10}
{\rm Res}_{\tau=0} \, \nu_{1,2} ^{(3)} = - \frac{12}{35} \neq 0.
\end{equation}
Thus, we have a nonzero residue at $\tau = 0$, which implies a logarithm in the solutions of (${\rm VE}_3$). Then, the
Galois group of (${\rm VE}_3$) is solvable, but not abelian. Again, the non-integrability of the Hamiltonian system (\ref{Hequ})
in this case follows from Theorem \ref{th3}.

\vspace{2ex}

\noindent
{\bf Step 6. Integrability of the case $n = 1, \quad \alpha_1 = \frac{1}{6}$.}

\vspace{2ex}

As we point out in the Introduction this integrable case has a multi-dimensional gene\-ra\-li\-zation.
Here we merely present the commuting integrals in our three-dimensional case for $A, B, C$ arbitrary.

With $\gamma = \alpha, \beta = 6 \alpha$ and after rescaling $t$ the Hamiltonian (\ref{1.1}) reads
\begin{equation}
\label{6.1}
H = \frac{1}{2} (p_1 ^2 + p_2 ^2 + p_3 ^2)
+\frac{1}{2} (\tilde{A} q_1 ^2 + \tilde{C} q_2 ^2 + \tilde{B} q_3 ^2) + ( q_1 ^2 +  q_2 ^2) q_3 + 2 q_3 ^3 ,
\end{equation}
where $\tilde{A} = A/\alpha, \tilde{B} = B/\alpha, \tilde{C} = C/\alpha$. Then the corresponding first integrals are
(see e.g. \cite{Eilbeck,Kostov})
\begin{eqnarray}
\label{6.2}
G_1 & =  & \frac{(q_2 p_1 - q_1 p_2)^2}{\tilde{A}- \tilde{C}} -q_1^4 - q_1^2 q_2^2 - 4 p_3 q_1 p_1 +
(\tilde{B} - 4 \tilde{A} + 4 q_3)(p_1^2 + \tilde{A} q_1^2),  \\
G_2 & = & -\frac{(q_2 p_1 - q_1 p_2)^2}{\tilde{A}- \tilde{C}} -q_2^4 - q_1^2 q_2^2 - 4 p_3 q_2 p_2 +
(\tilde{B} - 4 \tilde{C} + 4 q_3)(p_2^2 + \tilde{C} q_2^2). \nonumber
\end{eqnarray}

This completes the proof of Theorem 1.

$\hfill \blacksquare$

{\bf Acknowledgements.}

This work is partially supported by grant DN 02-5 of Bulgarian
Fund "Scientific Research".

\newpage

\appendix

\section{Some applications of the Morales-Ramis theory}

\subsection{Homogeneous potentials}

In this part we follow mainly the treatment in \cite{M}, which is
a generalization of a non-integrability theorem by Yoshida
\cite{Yoshida} for the case $n=2$ based on Ziglin's theorem.

Consider a Hamiltonian system with $n$ - degrees of freedom $(n
\geq 2)$ governed by a natural Hamiltonian
\begin{equation}
\label{s1} H = T + V = \frac{1}{2} (y_1^2 + \ldots + y_n^2) + V
(x_1, x_2, \ldots, x_n),
\end{equation}
where $V$ is a homogeneous function of degree $k \neq 0$.

Taking advantage of homogeneity, it is possible to obtain a
particular solution of the form
\begin{equation}
\label{s2} \Gamma: x_j = u (t) c_j, \quad y_j = \dot{u}(t) c_j,
\quad j = 1, \ldots, n ,
\end{equation}
where $u (t)$ is a solution of the hyperelliptic equation $
\dot{u}^2 = \frac{2}{k} (1-u^k) $ and $\mathbf{c}=(c_1, \ldots,
c_n)$ is a solution of the nonlinear system
\begin{equation}
\label{s4} \mathbf{c} =  V' (\mathbf{c}).
\end{equation}
Such solutions are called Darboux points. Then the (VE)
along $\Gamma$ are given by
\begin{equation}
\label{s5} \ddot{\mathbf{\xi}} = -u (t)^{k-2} V'' (\mathbf{c})
\mathbf{\xi}.
\end{equation}
Since $V'' (\mathbf{c})$ is diagonalizable, we can write
(\ref{s5}) as a direct sum of second order equations
\begin{equation}
\label{s6} \ddot{\xi}_j = -u (t)^{k-2} \lambda_j \xi_j, \quad j =
1, 2, \ldots, n,
\end{equation}
where $\lambda_j$ are the eigenvalues of $V'' (\mathbf{c})$
(usually called Yoshida coefficients).

In order to get (NVE) we rule out the equation corresponding to
$\lambda_n = k-1$
\begin{equation}
\label{s7} \ddot{\mathbf{\eta}} = -u (t)^{k-2} \mathrm{diag}
(\lambda_1, \ldots, \lambda_{n-1}) \mathbf{\eta},
\end{equation}
with $\mathbf{\eta} = (\xi_1, \ldots, \xi_{n-1})$.

Further, we change the independent variable by $x:= (u (t))^{k}$
and obtain a system of hypergeometric differential equations
\begin{equation}
\label{s8} x(1-x) \frac{d^2 \xi_j}{d x^2} + \left( \frac{k-1}{k} -
\frac{3k-2}{2 k} x \right) \frac{d \xi_j}{d x} +
\frac{\lambda_j}{2 k} \xi_j = 0, \quad j = 1, \ldots, n-1.
\end{equation}
It is clear that the identity component of the Galois group of
(\ref{s7}) is solvable (abelian) if, and only if, each one of the
identity components of the Galois groups of the hypergeometric
equations (\ref{s8}) is solvable (abelian).

Finally, exploring the Galois groups of the hypergeometric
equations (\ref{s8}) the following result is obtained.

\begin{thm}
\label{thA1} {\rm (Theorem 5.1 \cite{M})} If the Hamiltonian
system with the Hamiltonian (\ref{s1}) is completely integrable
with holomorphic or meromorphic first integrals, then each pair
$(k, \lambda_j)$ belongs to one of the following lists

\begin{tabular}{clcl}
 (1) & (k, s + s(s-1)k/2)                                            & (2) & $(2, \mathbb{C})$ \\
 (3) & $(-2, \mathbb{C})$                                            & (4) & $(-5, \frac{49}{40} - \frac{1}{40} ( \frac{10}{3} + 10 s )^2 )$ \\[0.5ex]
 (5) & $(-5, \frac{49}{40} - \frac{1}{40} ( 4 + 10 s )^2 )$          & (6) & $(-4, \frac{9}{8} - \frac{1}{8} ( \frac{4}{3} + 4 s )^2 )$ \\[0.5ex]
 (7) & $(-3, \frac{25}{24} - \frac{1}{24} ( 2 + 6 s )^2 )$           & (8) & $(-3, \frac{25}{24} - \frac{1}{24} ( \frac{3}{2} + 6 s )^2 )$ \\[0.5ex]
 (9) & $(-3, \frac{25}{24} - \frac{1}{24} ( \frac{6}{5} + 6 s )^2 )$ & (10)& $(-3, \frac{25}{24} - \frac{1}{24} ( \frac{12}{5} + 6 s )^2 )$ \\[0.5ex]
(11) & $(3, -\frac{1}{24} + \frac{1}{24} ( 2 + 6 s )^2 )$            & (12)& $(3, -\frac{1}{24} + \frac{1}{24} ( \frac{3}{2} + 6 s )^2 )$ \\[0.5ex]
 (13 & $(3, -\frac{1}{24} + \frac{1}{24} ( \frac{6}{5} + 6 s )^2 )$  & (14)& $(3, -\frac{1}{24} + \frac{1}{24} ( \frac{12}{5} + 6 s )^2 )$ \\[0.5ex]
 (15)& $(4, -\frac{1}{8} + \frac{1}{8} ( \frac{4}{3} + 4 s )^2 )$    & (16)& $(5, -\frac{9}{40} + \frac{1}{40} ( \frac{10}{3} + 10 s )^2 )$ \\[0.5ex]
 (17)& $(5, -\frac{9}{40} + \frac{1}{40} ( 4 + 10 s )^2 )$          & (18)&  $(k, \frac{1}{2}(\frac{k-1}{k} + s(s+1)k) )$
\end{tabular}

where $s$ is an arbitrary integer and $\mathbb{C}$ stands for an
arbitrary complex number.
\end{thm}

The above result is extended by Maciejewski, Przybilska and
Yoshida in several ways. We will use the following
\begin{thm}
\label{thA2} {\rm (Theorem 1.3 \cite{MPY})} Assume that the
Hamiltonian system defined by Hamiltonian (\ref{s1}) with a
homogeneous potential $V$ of degree $k \in \mathbb{Z} \setminus
\{0\}$ satisfies the following conditions:
\begin{enumerate}[label=(\arabic*)]
\item there exists a non-zero $\mathbf{c} \in \mathbb{C}^n$ such
that $\mathbf{c} =  V' (\mathbf{c})$, and \item matrix $V''
(\mathbf{c})$ is diagonalizable with eigenvalues $\lambda_1,
\ldots, \lambda_{n-1}, \lambda_n = k-1$; \item the system admits
an additional first integral $F$, which is meromorphic
 in a connected neighborhood $U$ of $\Gamma$.
\end{enumerate}
Then either
\begin{enumerate}[label=(\roman*)]
\item there exist $1 \leq r < n$ such that the pair $(k,
\lambda_r)$ belongs to the table in Theorem \ref{thA1}, or \item
there exist $1 \leq i < j < n$ such that
\begin{equation}
\label{s9} \frac{1}{2k} \sqrt{(k-2)^2 + 8 k \lambda_i} =
\frac{1}{2k} \sqrt{(k-2)^2 + 8 k \lambda_j} + l,
\end{equation}
for some $l \in \mathbb{Z}$.
\end{enumerate}
\end{thm}

\subsection{Necessary conditions for integrability of Hamiltonian systems which have (NVE) of Lam\'e type}

Here we recall some facts concerning the integrability of
Hamiltonian systems with two degrees of freedom, an invariant
plane and which (NVE) are of Lam\'e type. More details can be
found in \cite{MSim,M}. In our case the (NVE) splits into two
equations of Lam\'e type, and therefore, these arguments can be
applied.

Classically the Lam\'e equation is written in the form
\begin{equation}
\label{Lame} \ddot{\xi} - [n(n+1) \wp(t) + B] \xi = 0,
\end{equation}
where $\wp (t)$ is the Weierstrass function with invariants $g_2$
and $g_3$, satisfying $\dot{v}^2 = 4 v^3 - g_2 v - g_3$ with
$\Delta = g_2 ^3 - 27 g_3 ^2 \neq 0$.

The known (mutually exclusive) cases of closed form solutions of
(\ref{Lame}) are:

(i) The Lam\'e and Hermite solutions. In this case $n \in
\mathbb{Z}$ and $g_2, g_3, B$ are arbitrary parameters;

(ii) The Brioschi-Halphen-Crowford solutions. Here $m:= n + 1/2
\in \mathbb{N}$ and the parameters $g_2, g_3, B$ must satisfy an
algebraic equation.

(iii) The Baldassarri solutions. Now $n + 1/2 \in \frac{1}{3}
\mathbb{Z} \cup \frac{1}{4} \mathbb{Z} \cup \frac{1}{5} \mathbb{Z}
\setminus \mathbb{Z}$ with additional algebraic relations between
the other parameters.

Note that in the case (i) the identity component of the Galois
group $G^0$ is of the form $\begin{pmatrix}
1 & 0 \\
\nu & 1
\end{pmatrix}$
 and in the cases (ii) and (iii) $G^0 = id$ ($G$ is finite).
And these are the all cases when the Lam\'e equation is
integrable.

Now consider a natural two degrees of freedom Hamiltonian
\begin{equation}
\label{ham2} H = \frac{1}{2} (p_1 ^2 + p_2 ^2) + V (q_1, q_2),
\end{equation}
$q_j (t) \in \mathbb{C}, p_j (t) = \dot{q}_j, j=1, 2$. We assume
that there exists a family of solutions of the form
$$
\Gamma_h : q_2 = p_2 = 0, \quad q_1 = q_1 (t, h), \quad  p_1 (t,
h) = \dot{q}_1 (t, h)
$$
and $q_1 (t, h)$ is a solution of
$$
\frac{1}{2} \dot{q}_1 ^2 + \varphi (q_1) = h, \quad h \in
\mathbb{R} .
$$
The (NVE) along $\Gamma_h$ is
\begin{equation}
\label{nve} \ddot{\xi} - f (t, h) \xi = 0 ,
\end{equation}
where $f (t, h) = f (q_1 (t, h))$ is such that (\ref{nve}) is of
type (\ref{Lame}).

In \cite{MSim,M} the type of the potentials $V$ with this property
are obtained as well as the necessary conditions for the
integrability of the Hamiltonian systems with the Hamiltonian
(\ref{ham2}). In order to formulate the result we need certain
additional quantities.

Since $f (t, h)$ depends linearly on $\wp (t)$, then $\dot{f}^2$
is a cubic polynomial in $f$, depending also in $h$, namely
\begin{equation}
\label{apl} \dot{f}^2 : = P (f, h) = P_1 (f) + h P_2 (f) .
\end{equation}
The following coefficients are introduced
\begin{equation}
\label{coeff} P (f, h) = (a_1 + h a_2) f^3 + (b_1 + h b_2) f^2 +
(c_1 + h c_2) f + (d_1 + h d_2) .
\end{equation}
Now we are ready to give the corresponding result. Note that the
following Theorem gives necessary conditions only from the
analysis of the first variational equation.
\begin{thm}
\label{thB1} {\rm(Theorem 6.2 \cite{M})}. Assume that a natural
Hamiltonian system has (NVE) of Lam\'e type, associated to the
family of solutions $\Gamma_h$, lying on the plane $q_2 = 0$ and
parameterized by the energy $h$. Then, a necessary conditions for
integrability is that the related polynomials $P_1$ and $P_2$
satisfy $a_2 = 0$, and one of the following conditions holds:

\vspace{2ex}

\noindent {\rm I}. $a_1 = \frac{4}{n (n+1)}$ for some $n \in
\mathbb{N}$;

\vspace{2ex}

\noindent {\rm II}. $a_1 = \frac{16}{4 m^2 - 1}$ for some $m \in
\mathbb{N}$. Then, assuming the conjecture above is true, one
should have $b_2 = 0$ and we should be in one of the following
cases:

\vspace{1ex}

${\rm II}_1$  $m = 1$ and $b_1 = 0$,

\vspace{1ex}

${\rm II}_2$  $m = 2$ and $c_2 = 0, \, 16 a_1 c_1 + 3 b_1 ^2 = 0$,

\vspace{1ex}

${\rm II}_3$ $m = 3$ and $16 a_1 d_2 + 11 b_1 c_2 = 0, \, 1024 a_1
^2 d_1 + 704 a_1 b_1 c_1 + 45 b_1 ^3 = 0$,

\vspace{1ex}

${\rm II}_m$ $m > 3$. Then, we should have $b_1 = 0$ and,
furthermore, either $c_1 = c_2 = 0$ if $m$ is congruent with $1,
2, 4$ or $5$ modulo $6$, or $d_1 = d_2 = 0$ if $m$ is odd;

\vspace{2ex}

\noindent {\rm III}.  $a_1 = \frac{4}{n (n+1)}$ with $n + 1/2 \in
\frac{1}{3} \mathbb{Z} \cup \frac{1}{4} \mathbb{Z} \cup
\frac{1}{5} \mathbb{Z} \setminus \mathbb{Z}$, $b_2 = 0$ and either
$c_2 = 0, b_1 ^2 - 3 a_1 c_1 = 0$ or $c_2 b_1 - 3 a_1 d_2 = 0, 2
b_1 ^3  - 9 a_1 b_1 c_1 + 27 a_1 ^2 d_1 = 0$.
\end{thm}
It is clear that the condition I. in the above Theorem gives the
Lam\'e and Hermite solutions (i), the condition II.-- the
Brioschi-Halphen-Crowford solutions (ii), and the condition III.
-- the Baldassarri solutions (iii).


\begin{thebibliography}{99}

\bibitem{HH}
M. H\'{e}non, C. Heiles,
The Applicability of the Third Integral of Motion: Some Numerical Experiments,
The Astron. J. {\bf 69} 7-79 (1964)


\bibitem{Wenlei}
Wenlei Li, Shaoyun Shi,
Non-integrability of H\'{e}non-Heiles system,
Celest. Mech Dyn Astr, 2010, DOI 10.1007/s10569-010-9315-1

\bibitem{Fer1}
S. Ferrer, J. Palaci\'{a}n, J.F. San Juan, A. Viartola, P. Yanguas,
On the H\'{e}non and Heiles system in three dimensions: The role of the axial symmetry,
Phys. lett. A {\bf 228} 255-260 (1997)

\bibitem{Fer2}
S. Ferrer, M. Lara, J. Palaci\'{a}n, J.F. San Juan, A. Viartola, P. Yanguas,
The H\'{e}non and Heiles problem in three dimensions. I. Periodic orbits near the origin,
Int. J. of Bifurcation and Chaos, {\bf 8}, No. 6, 1199-1213 (1998)

\bibitem{Eilbeck}
J.C. Eilbeck, V. Z. Enol’skii, V.B. Kuznetsov C and D.V. Leykin,
Linear r-matrix algebra for systems separable in parabolic coordinates,
Physics Letters A {\bf 180} (1993) 208—214


\bibitem{Kostov}
N.A. Kostov, V.S. Gerdjikov, V. Miok,
Exact solutions for a class of integrable H\'{e}non-Heiles-type system,
J. Math. Phys. {\bf 51} 022702 (2010), doi.org/10.1063/1.3300310


\bibitem{Zeng} Yunbo Zeng,
A Hierarchy of Multidimensional H\'{e}non-Heiles Systems,
Acta Mathematica Sinica, English Series {\bf 16} No. 3, 527-534 (2000)


\bibitem{Fakkousy}
I. Fakkousy, J. Kharbach, W. Chater, M. Benkhali, A. Rezzouk, M. Quazzani-Jamil,
Liouvillian integrability of the three-dimensional generalized H\'{e}non-Heiles Hamiltonian,
Eur. Phys. J. Plus, 2020, 135:612 doi.org/10.1140/epjp/s13360-020-000625-z


\bibitem{Ito}
Ito H.,
Non-integrability of H\'{e}non-Heiles system and a theorem of Ziglin,
J. Kodai. Math. {\bf 8} 120-138 (1985)

\bibitem {M}
J. Morales Ruiz, Differential Galois Theory and Non integrability
of Hamiltonian Systems, Prog. in Math., v. 179, Birkh\"auser,
1999.

\bibitem{MRS1}
J. Morales-Ruiz, J-P. Ramis, and C. Sim\'o,
Integrability of Hamiltonian systems and differential Galois groups of higher variational equations,
Annales scientifiques de l'\'{E}cole normale sup\'{e}rieure 40 (2007) 845--884.

\bibitem{MR2} Morales-Ruiz J., Ramis J-P.,
Integrability of Dynamical systems through Differential Galois
Theory: practical guide. Contemporary Math 2010; 509.

\bibitem{SvP} Van der Put M., Singer M.,
Galois Theory of Linear Differential Equations.
In: Grundlehren der Mathematischen Wissenschaften, vol. 328. Berlin: Springer; 2003.

\bibitem{Z1} Ziglin S.L.,
Branching of solutions and non-existence of first integrals in
Hamiltonian mechanics:
Funct. Anal. Appl. I {\bf 16} (1982) 181-189.


\bibitem{Z2} Ziglin S.L.,
Branching of solutions and non-existence of first integrals in
Hamiltonian mechanics: Funct. Anal. Appl. II {\bf 17} (1983) 6-17.


\bibitem{Yoshida}
H. Yoshida, A criterion for non-existence of an additional integral in hamiltonian systems with a homogeneous potential,
Physica D 29, 1987, 128-142.


\bibitem{Hie}
J. Hietarinta,
Direct methods for the search of the second invariant,
Phys. Rep. {\bf 147} (1987), 87-154.

\bibitem{MacPrzy} A. Maciejewski, M. Przybilska,
All meromorphically integrable $2D$ Hamiltonian systems with homogeneous potential of degree 3,
Phys. Lett. A, {\bf 327}, (2004), 461-473.


\bibitem{MPY} A. Maciejewski, M. Przybilska, H. Yoshida,
Necessary conditions for the existence of additional first integrals for Hamiltonian systems with homogeneous potential,
Nonlinearity, {\bf 25} (2012) 255-277.


\bibitem{WW}
Wittaker E., Watson E.,
A cource of Modern Analysis, Cambridge University Press, Cambridge, UK, 1989


\bibitem{MSim}
Morales-Ruiz J, Sim\'o C.
Non-integrability Criteria for Hamiltonians in the case of Lam\'e Normal Variational Equations,
J Diff Eq 1996; 129: 111-135.


\end{thebibliography}
\end{document}